\documentclass[12pt,epsfig]{article}

\renewcommand{\baselinestretch}{1.2}

\usepackage[dvips]{graphics}
\usepackage{rotating}
\usepackage{epsfig}
\def\lesssim{\mathrel{\hbox{\rlap{\hbox{\lower5pt\hbox{$\sim$}}}\hbox{$<$}}}}
\def\gtrsim{\mathrel{\hbox{\rlap{\hbox{\lower5pt\hbox{$\sim$}}}\hbox{$>$}}}}

\newcommand{\ntrl}[1]{\chi^0_#1}

\newcommand{\chp}[1]{\chi^+_#1}

\newcommand{\sbot}[1]{\tilde{b}_#1}
\newcommand{\sstop}[1]{\tilde{t}_#1}

\newcommand{\sbotc}[1]{\tilde{b}_#1^*}
\newcommand{\sstopc}[1]{\tilde{t}_#1^*}
\def\gluino{\tilde{g}}

\def\bbar{\bar{b}}       %
\def\tbar{\bar{t}}       %

\newcommand{\mntrl}[1]{m_{\chi^0_#1}}      %
\newcommand{\mchpm}[1]{m_{\chi^\pm_#1}}

\newcommand{\msbot}[1]{m_{\tilde{b}_#1}}
\newcommand{\mstop}[1]{m_{\tilde{t}_#1}}
\def\mgluino{m_{\tilde{g}}}

\def\tanbeta{\tan\beta}

\def\pt{p_T}

\def\ptjet{p_T^{jet}}

\def\beq{\begin{equation}}   %
\def\eeq{\end{equation}}   %

\begin{document}

\begin{flushright}
   {\bf HRI-P07-10-001 \\
   HRI-RECAPP-07-14}
\end{flushright}

\vskip 30pt

\begin{center}
{\large \bf  Associated Higgs Production in CP-violating supersymmetry:
probing the `open hole' at the Large Hadron Collider}\\
\vskip 20pt
{Priyotosh Bandyopadhyay$^{a}$\footnote{priyotosh@mri.ernet.in},
Amitava Datta$^b$\footnote{adatta@juphys.ernet.in},
AseshKrishna Datta$^a$\footnote{asesh@mri.ernet.in} \\ and 
Biswarup Mukhopadhyaya$^a$\footnote{biswarup@mri.ernet.in}}  \\
\vskip 20pt
{$^a$ Regional Centre for Accelerator-based Particle Physics  \\
Harish-Chandra Research Institute  \\
Chhatnag Road, Jhunsi, Allahabad, India 211019 }\\
\vskip 5pt
{$^b$ Department of Physics \\
Jadavpur University, Kolkata, India 700032}
\end{center}

\vskip 65pt

\abstract{A benchmark $CP$-violating supersymmetric scenario 
(known in the literature as `CPX-scenario')
is studied in the context of  the Large Hadron Collider (LHC). It is shown
that the LHC, with low to moderate
accumulated luminosity, will be able to probe the existing `hole' in the 
$m_{h_1}$-$\tan\beta$ plane, which cannot be ruled out by the Large Electron Positron 
Collider data. This can be done 
through associated production of Higgs bosons
with top quark and top squark pairs leading to the signal
\emph{dilepton + $\leq{5}$ jets (including $3$ b-jets) + missing ${p_T}$}.
Efficient discrimination of such a $CP$-violating supersymmetric
scenario from other contending ones is also possible at 
the LHC with a moderate volume of data.}

\newpage

\section{Introduction}

One of the main motivations for suggesting supersymmetry (SUSY) is to remove the fine-tuning
problem in the Higgs sector of the standard model.
The condition of holomorphicity of the superpotential requires two Higgs doublets
in the minimal SUSY extension of the standard model (SM). There the Higgs sector has a 
larger particle content than the SM, and the physical states in this sector
comprise two neutral scalars, one pseudoscalar and one charged Higgs boson.
Finding the signatures of these  scalars is thus inseparably linked with
the search for SUSY at the upcoming Large Hadron Collider (LHC).

Prior to the LHC several Higgs search experiments have yielded negative results. The strongest
lower bound on the smallest Higgs mass ($m_h$)  from the Large Electron Positron Collider (LEP) 
is $m_h > $ 114.4 GeV \cite{Barate:2003sz,Schael:2006cr}. This limit is valid for a SM like Higgs as well as for the lightest
neutral Higgs boson in the minimal supersymmetric standard model (MSSM) in the decoupling limit
i.e. the limit in which the masses of all other scalars in the Higgs sector become very large.
Although smaller values of $m_h$ are allowed away from the decoupling limit, the lower bound on its mass is approximately the $Z$-mass. However, when the Higgs sector inherits some
$CP$-violating phase through radiative corrections \cite{Pilaftsis:1998pe,Pilaftsis:1998dd}, the above limit ceases to be valid. Our discussion
is centred around such situations.

It is well-known by now that lower bound on the mass of the lightest Higgs boson of 
the $CP$-conserving MSSM 
 from LEP \cite{Schael:2006cr} can be drastically reduced or even may entirely vanish if non-zero 
$CP$-violating phases are allowed \cite{Carena:2000ks}. This can happen through 
radiative corrections to the Higgs potential, whereby the phases, if any, 
of the Higgsino mass parameter $\mu$ 
and the trilinear soft SUSY breaking parameter $A$ enter into the picture. As a result 
of the $CP$-violating phase, the neutral spinless states are no more of definite parity,
and their couplings to gauge bosons as well as fermions are thus modified, depending on
the magnitude of the phases. Thus there are three neutral states $h_i$ ($i$=1,2,3);
the collider search limits for all of them are modified since the squared amplitudes
for production via $WW$, $ZZ$ and $q\bar{q}$ couplings for all of them now consist of
more than one terms. Mutual cancellation among such terms can take place in certain 
regions of the parameter space, thus resulting in reduced production rates and
consequent weakening of mass limits at collider experiments.

For example, in the context of a
benchmark $CP$-violating scenario (often called the CPX scenario in the
literature \cite{Carena:2000ks}), 
it has been found that $m_{h_1}$ as low as  50 GeV or even smaller, cannot be ruled out by the 
final LEP data for low and moderate values of  $\tan\beta$, where $h_1$ is the lightest
neutral Higgs, and $\tan\beta$ is the 
ratio of the vacuum expectation values of the two Higgs doublets. 
In other words, a `hole' is found to exist in the $m_{h_1}$-$\tanbeta$ parameter space 
covered by the LEP searches, the underlying reason being the reduction
in  the coupling $ZZh_1$ due to the $CP$-violating phase(s), as mentioned above. 
Moreover, complementary channels such as
$e^+ e^- \to h_1 h_2$,  suffer from coupling as well as phase-space suppression 
within this `hole', thus making it inaccessible to  LEP  searches.
The existence of this hole has been confirmed by the analyses of 
the LEP data by different experimental groups \cite{Schael:2006cr,Carena:2000ks,Bechtle:2006iw}, although
people are not unanimous about the exact span of the hole.

The next natural step is to assess the prospect of closing the hole at Tevatron 
Run II or the LHC. 
The existing analysis on this \cite{Carena:2002bb}, however, focuses on the discovery channels based on
the conventional Higgs production and decay 
mechanisms  employed in the context of the SM.  It has been noted 
that although the hadron colliders can probe most of the parameter space of the 
CPX scenario and can indeed go beyond some regions of the parameter space scanned
by the  LEP searches, the lightest Higgs boson within the aforementioned hole may 
still escape detection. This is because not only $ZZh_1$ but also the $WWh_1$ and 
$t\bar{t}h_1$ couplings tend to be very small within this hole.  
On the other hand, the relatively heavy neutral Higgs bosons $h_{2,3}$ couple to $W$, $Z$ 
and $t$ favourably, but they can decay in non-standard channels, thus requiring
a modification in search strategies.
The work \cite{Accomando:2006ga} which has compiled possible signals of the
CPX scenario at the LHC is also restricted  to the
production of $h_i$ ($i$=1,2,3) bosons in SM-like channels. However, it
looked into more decay channels of the $h_i$ bosons thus produced.
It has been henceforth concluded that parts of
the holes in the $M_H^+$-$\tan\beta$ or the $m_{h_1}$-$\tan\beta$
parameter space can be plugged, although considerable portions of the hole, 
especially for low $\tan\beta$, may escape detection
at the  LHC even after accumulating 300 fb$^{-1}$ of integrated luminosity.

Thus it is important to look for other production channels for the scalars in the 
CPX region, especially by making use of the couplings of $h_1$ with the sparticles. 
It is gratifying to note in this context that 
the $\tilde{t}_{1}\tilde{t}^{*}_{1}h_1$ coupling, where $\sstop1$ is the lighter top squark, 
indeed leads to such a discovery channel, in cases where the $t$-$\bar{t}$-$h_1$ 
and  $W$-$W$-$h_1$, $Z$-$Z$-$h_1$  couplings are highly suppressed. In 
fact it has been noted that in a general $CP$-violating MSSM, 
the cross section of $\sstop1\sstopc1h_1$ production 
could be dramatically larger than that obtained by switching off the $CP$-violating 
phases \cite{Li:2006hq}. Since the trilinear SUSY breaking parameter $A_t$ 
is necessarily large in the CPX scenario, 
$\sstop1$ tends to be relatively light and may be produced at 
the LHC with large cross section. As a bonus, both $h_2$ and $h_3$ also couple 
favourably to the $t\bar{t}$ pair and can add modestly to the signal although by 
themselves they fail to produce a statistically significant signal. In this work we investigate
the implications of these couplings at the LHC, by concentrating on a specific signal 
arising from the associated production of the neutral Higgs bosons with a 
top-pair or a pair of lighter stop squarks.

Our task, however, does not end here. While we wish to extract information
on the neutral Higgs sector in the CPX scenario, 
other SUSY processes driven by other particles in the spectrum 
may yield the same final state. To make sure that one is indeed looking at the Higgs
sector, one needs to isolate the Higgs-induced channels, and find event
selection criteria to not only reduce the SM backgrounds but also ensure that
the canonical SUSY channels do not overwhelm the Higgs signatures.
In our analysis, we first introduce suitable criteria which will suppress the SM 
background compared to the total SUSY contribution in CPX. Next, we  
suggest additional discriminators for further filtering out the contributions 
of  the lightest Higgs ($h_1$) from other SUSY channels. We finally show that if 
nature prefers  the SM alone with $m_h \geq$ 114.4 GeV, or, 
alternatively, $CP$-conserving SUSY, the proposed signal would indeed be much 
smaller if our selection criteria are imposed.

The paper is organised as follows. In Section 2 we discuss the basic
inputs of the CPX scenario, the resulting mass spectrum and other
features they lead to. All of our subsequent numerical analysis would be
in this framework where we also use the alternative expression CPV-SUSY to 
mean the CPX-scenario.  In section 3 we set out to define the proposed signal,
devise the event selection criteria to reduce both SM and residual SUSY
backgrounds and fake events, and present the final numerical results.
We summarise and conclude in section 4.

\section{The CPX  Model: values of various parameters} 

As indicated in the Introduction, we adopt the so called CPX 
scenario in which the LEP analyses have been
performed. It has been observed
\cite{Pilaftsis:1998pe, Pilaftsis:1998dd}
that the $CP$-violating quantum effects on the Higgs potential is proportional to
$Im(\mu A_t)/M^2_{SUSY}$, where $A_t$ is the trilinear soft SUSY breaking parameter occurring
in the top squark mass matrix, and $M_{SUSY}$ is the characteristic SUSY breaking scale,
being of the order of the third generation squark masses. 
With this in mind, a benchmark scenario known as CPX was proposed \cite{Carena:2000ks} and
its consequences were studied
[\cite{Demir:1999hj}--\cite{Ham:2002ps}]
in some of which steps are suggested for closing the aforementioned `hole' \cite{Akeroyd:2003jp,Ghosh:2004cc,Ghosh:2004wr}.
In this scenario, the effects of $CP$-violation are maximized.
The corresponding inputs that we adopt here 
are compatible with the ``hole'' left out in the analysis.

\hskip 47pt $m_{\tilde{t}} = m_{\tilde{b}} = m_{\tilde {\tau}} = M_{SUSY} = 500$ GeV, 
 $\qquad \mu = 4 M_{SUSY} = 2$ TeV 

\hskip 50pt $|A_t| = |A_b| = 2 M_{SUSY} = 1$ TeV, $\quad arg(A_{t,b}) = 90^\circ$ 

\hskip 50pt $|\mgluino|= 1$ TeV, $\quad arg(\mgluino) = 90^\circ$

\hskip 50pt  $M_2 = 2 M_1 = 200$ GeV,~~~~$\tan\beta= 5 - 10$ \\

\noindent
where the only departure from reference \cite{Carena:2002bb} lies in a small tweaking 
in the mass ratio of the $U(1)$ and $SU(2)$ gaugino masses 
$M_1$ and $M_2$, aimed at ensuring gaugino mass unification at high scale. 
It has been checked that this difference does not affect the Higgs production
or the decay rates \cite{PvtApost}.
The presence of a relatively large $A_t$ ensures that one of the top squarks will
be relatively light. The value of the top quark mass has been taken to be 175 GeV\footnote{The frequent shift in the central
  value of $m_t$, coming from Tevatron measurements, causes the size of the
  hole to change, although its location remains the same. 
However, there is little point in worrying about this uncertainty, since the 
very quantum corrections which are at the root of all $CP$-violating 
effects in the Higgs sector are prone to similar, if not greater, 
theoretical uncertainties.}.

It is to be noted that the first two generation sfermion masses must be kept
sufficiently heavy so that the stringent experimental bound (for example, 
the electric dipole moment of the neutron) is satisfied. Here we have not 
considered possible ways of bypassing such bounds, and set the masses of the first 
two sfermion families at 10 TeV. Thus our analysis is based on
the mass spectrum showed in Table 1.

\begin{table}[h]
\begin{center}
\renewcommand{\arraystretch}{1.4}
\begin{tabular}{|c|c|c|c|c|c|c|c|c|c|c|}
\hline
$m_{h_1}$ & $m_{h_2}$ & $m_{h_3}$ & $\mstop1$ & $\mstop2$ & $\msbot1$
& $\msbot2$ & $\mntrl1$ & $\mntrl2$ & $\mchpm1$ \\
\hline
48.9 & 103.3 & 135.7 & 322.0 & 664.0 & 476 & 527 & 99.6 & 198.4 & 198.4 \\
\hline
\end{tabular}
\caption{Physical masses (in GeV) of neutral Higgs bosons, squarks and lighter
  gauginos in the CPX scenario with $\tan\beta$=5 and  $m_{H^\pm}$=$130$ GeV.}
\end{center}
\end{table}

\noindent
The specific choice of $m_{H^\pm}$ is made to obtain the mass of the lightest Higgs boson 
within the LEP-hole in $m_{h_1}$-$\tanbeta$ space. It should be noted that such a choice 
makes the remaining two neutral Higgs bosons not so heavy either. This kind of
a situation has a special implication in CPV-MSSM, namely, all the neutral
Higgs bosons can be produced in association with a $\sstop1$ pair. Such production
is kinematically suppressed in the $CP$-conserving case due to the lower bound
on $m_h$.

 The CPX set of parameters listed above constitutes our benchmark point number 1
(BP1) in the detailed analysis to be undertaken in the next section. We list
at the end of that section the final results corresponding to six more benchmark points within the
hole unprobed by current data. These points are denoted by BP2 - BP7.

\section{Signals at the LHC}

Since, in CPX-SUSY the $VVh_1$ ($V$=$W,Z$) and  $t\bar{t}h_1$ interactions are suppressed
for the lightest neutral scalar($h_1$), we shall have to think of some alternative
associate production mechanism at the LHC. One possibility is to consider
the associated production of $h_1$ with a pair of lighter stops. The large value of
$A_t$ is encouraging in this respect. In addition, since the point CPX yields a
not-so-high value of the lighter stop mass,  this production mechanism is 
kinematically quite viable.

The cross sections for different supersymmetric associated 
production processes are computed with {\tt CalcHEP} \cite{Pukhov:2004ca} (interfaced with the program {\tt CPSuperH}
\cite{Ellis:2006eh,Lee:2003nta})
and listed in Table 2. As one can see, while a substantial production rate
is predicted for  $h_1$ associated with a pair of $\sstop1$, 
the corresponding cross sections for $h_2$ and $h_3$ 
are smaller by two orders of magnitude. This is not only because of 
phase space suppression for the latter at the CPX point, but also
due to the conspiracy of a number of terms in the effective interaction
involved. Table 2 also reveals a complementary feature in Higgs production in association with a pair of 
top quarks, the underlying reason being again the multitude of
terms that enters into the squared amplitudes, and the provision of their
mutual cancellation in the CPX scenario. Thus we can identify,
for the given set of input parameters, $\sstop1 \sstopc1 h_1$ and $t \bar{t} 
h_{2,3}$ as the production processes that can be potentially useful in 
closing the hole in the parameter space. 

Also indicated in Table 2 is the gluino pair production cross section in the CPX
scenario for $\mgluino=$1 TeV which is a CPX input indicated earlier in this section. 
Later in this section, we shall explain how this process could affect our signal. 

 \vskip 10pt
 \renewcommand{\arraystretch}{1.4}
\begin{table}[h]
 \begin{center}
\begin{tabular}{|c|c|c||c|c|c||c|}
 \hline
 $\sigma_{\sstop1 \sstopc1 h_1}$  &  $\sigma_{\sstop1 \sstopc1 h_2}$  & $\sigma_{\sstop1 \sstopc1 h_3}$  &  
$\sigma_{t \bar{t} h_1}$  &  $\sigma_{t \bar{t} h_2}$  & $\sigma_{t \bar{t}
   h_3}$ & $\sigma_{ \tilde{g}\tilde{g}}$ \\
\hline
 440 & 6 & 4 & 8 & 198 & 135 & 134 \\
 \hline
\end{tabular}
 \caption{Production cross sections (in fb) at lowest-order 
computed with {\tt CalcHEP} interfaced with CPsuperH
 for different signal processes at the LHC in the
CPX scenario and for the spectrum of Table 1. 
 {\tt CTEQ6L} parton distribution functions are used and
 the renormalization/factorization scale is set to  $\sqrt{\hat{s}}$.}
\end{center}
 \end{table}

 The branching fractions of the lighter scalar top and the lightest neutral Higgs boson plays a
crucial role in selecting the viable modes in which the signal for CPV-SUSY can be looked for.
 In Table 3 we present the relevant branching fractions, keeping in mind that new final states
emerge whenever the branching fraction for a heavier neutral scalar decaying into two
 lighter ones is of sizable magnitude. In any case, it is interesting to note that 
not only the lightest Higgs $h_1$ but also $h_2$ and $h_3$ could play significant roles in 
 signals of the Higgs sector in the CPX scenario, given the possibility of all of them being
rather light.

 \renewcommand{\arraystretch}{1.4}
\begin{table}[h]
 \begin{center}
\begin{tabular}{|c|c||c|c|c||c|}
 \hline
Br($\sstop1 \to b \chp1$) &  Br($\sstop1 \to t \ntrl1)$  & Br($h_1 \to b \bar{b}$)  & 
 Br($h_2 \to h_1 h_1 $) &Br($h_3 \to h_1 h_1$) & Br($\tilde{g} \to t \tilde{t^{*}_1}$)  \\
 \hline
0.81 & 0.19 & 0.91 &  0.71& 0.82 &0.16 \\
 \hline
\end{tabular}
 \end{center}
\caption{Branching fractions for lighter top squark, the neutral Higgs bosons
   and gluino in the CPX scenario.}
\end{table}

Before we enter into the discussion of our specifically chosen signal, let us mention that,
in this study, {\tt CalcHEP} (interfaced to the program {\tt CPSuperH}) has also been used for generating parton-level events for the relevant processes. The standard {\tt CalcHEP-PYTHIA} interface \cite{alex}, which uses the SLHA
 interface \cite{Skands:2003cj} was then used to pass the {\tt CalcHEP}-generated 
events to {\tt PYTHIA} \cite{Sjostrand:2001yu}. Further, all relevant decay-information are generated with 
 {\tt CalcHEP} and are passed to {\tt PYTHIA} through the same interface.  All these 
 are required since there is no public implementation of CPV-MSSM in {\tt PYTHIA}. Subsequent 
decays of the produced particles, hadronization and the collider analyses are done with {\tt PYTHIA (version 4.610)}. 

We used {\tt CTEQ6L} parton distribution function (PDF) \cite{Lai:1999wy,Pumplin:2002vw}. 
In {\tt CalcHEP} we opted for the lowest order $\alpha_s$ evaluation, which is appropriate for a lowest order PDF 
like {\tt CTEQ6L}. The renormalization/factorization scale in {\tt CalcHEP} is set at $\sqrt{\hat{s}}$. This choice of scale results in a somewhat conservative estimate for the event rates.

 As discussed earlier, the processes of primary importance for the present 
study are $pp \to \tilde{t}_1\tilde{t}^*_1 h_{1}$ and $pp \to t\bar{t} h_{2,3}$. At the 
parton level, the lightest Higgs  and both top quarks (or top squarks) dominantly decay to 
$b$ quarks. For our signal, the associated $W$'s (or charginos) 
produced in  the decay of $t$ (or $\sstop1$) are required to decay into leptons with known or 
calculable branching ratios.
These decays lead to a final state with \emph{four} $b$-quarks along with
other SM particles.  In addition, the large branching ratios for $h_{(2,3)}
\to h_1 h_1$ can make the modest contributions from the $t\bar{t}h_{(2,3)}$ 
particularly rich in final state $b$'s, which, with a finite $b$-tagging 
efficiency, can provide a combinatoric factor of advantage to us.

 However, although $h_1$ decays dominantly into $b\bar{b}$, our simulation reveals that in a fairly large 
fraction of events both the $b$-quarks do not lead to sufficiently hard jets
with reasonable $b$-tagging efficiency. This is because of the lightness of $h_1$ in this scenario.
To illustrate this, we present in Figure 1 the ordered $p_T$ distributions for the four parton-level $b$-quarks in the
signal from $\sstop1 \sstopc1 h_1$. It is clear from this figure that the 
$b$-quark with the lowest $p_T$ in a given event is
often below 40 GeV or thereabout, which could have ensured a 
moderate tagging efficiency ($\geq$ 50\%). This forces us to
settle for three tagged $b$-jets in the final state, and look for 
$$
3~\emph{tagged $b$-jets + dilepton + other untagged jets + missing $\pt$}. 
$$
Later in this section we will
demonstrate that this feature is retained under a realistic situation, i.e.
on inclusion of hadronization.

\begin{figure}[hbt]
\begin{center}
\vspace*{-4.2cm}
\centerline{\epsfig{file=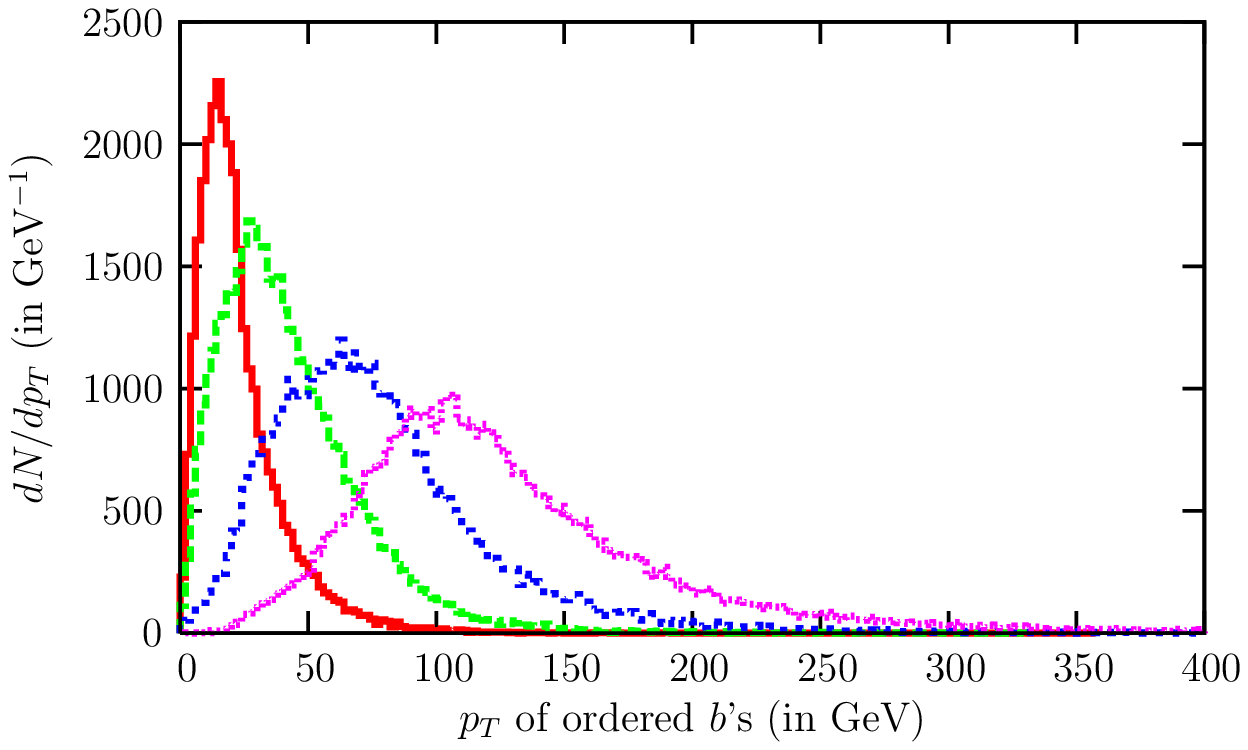,width=15.0 cm,height=30.0cm,angle=0}}
\vspace*{-17.5cm}
\caption{Ordered $p_T$ distributions for all four parton level $b$-jets arising from the 
decays of $\sstop1$, $\sstopc1$ and $h_1$ in 
  $\tilde{t_1}\tilde{t^{*}_{1}}h_1$ production.}
\end{center}
\label{fig11}
\vspace*{-1.0cm}
\end{figure}

We have used {\tt PYCELL}, the toy calorimeter simulation provided in
{\tt PYTHIA}, with the following criteria:
\begin{itemize}
  \item the calorimeter coverage is $\rm |\eta| < 4.5$ and the segmentation is
given by $\rm\Delta\eta\times\Delta\phi= 0.09 \times 0.09 $ which resembles
        a generic LHC detector
  \item a cone algorithm with
        $\Delta R = \sqrt{\Delta\eta^{2}+\Delta\phi^{2}} = 0.5$
        has been used for jet finding
  \item $ p_{T,min}^{jet} = 30$ GeV and jets are ordered in $p_{T}$
  \item leptons ($\rm \ell=e,~\mu$) are selected with
        $p_T \ge 30$ GeV and $\rm |\eta| \le 2.5$
  \item no jet should match with a hard lepton in the event
  
\end{itemize}
In addition, the following set of basic (standard) kinematic cuts is incorporated throughout our analysis:
\begin{center}
$p_T^{j_{1,2}}\geq{50}$ GeV 
$p_T^{j_3} \geq 40$ GeV $\quad$ $|\eta| _{j,\ell}\leq 2.5$  \\
$\Delta R_{\ell j} \geq 0.4$   $\quad$  $\Delta R_{\ell\ell}\geq 0.2$ 
\end{center}

\noindent
where $\Delta R_{\ell j}$ and $\Delta R_{\ell \ell}$ measure the lepton-jet and lepton-lepton isolations
respectively, with $\Delta R = \sqrt{\Delta \eta ^2 + \Delta \phi ^2}$, $\Delta\eta$ being the
pseudo-rapidity difference and $\Delta\phi$ being the difference in azimuthal angle for the
adjacent leptons and/or jets. Since efficient identification of  the leptons  is crucial for our study,
we required, on top of above set of cuts, that hadronic activity within a cone of 
$\Delta R=0.2$ between two isolated leptons should be minimum with $\ptjet<10$ GeV in the
specified cone. Throughout the analysis  we have assumed that a $b$-jet with $\ptjet> 40$ GeV 
can be tagged with 50\% probability. In addition, as we shall see below, some further kinematic cuts 
are  necessary to make the proposed signal stand out.

Below the contributions to the final state 
from different scenarios are discussed: 

\begin{itemize}
\item Contributions coming from the CPV-SUSY scenario and comprised
of $pp \to \sstop1 \sstopc1 h_1, t \tbar h_{2,3}$ and $pp \to \gluino \gluino$ where $m_{h_1}$ could
escape the LEP bound and can be as light as 50 GeV for low to moderate $\tan\beta$.
\item If nature is supersymmetric but conserves $CP$ (CPC-SUSY), contributions could dominantly
come from $pp \to t\tbar h$ and $\gluino \gluino$, where the appropriate LEP 
bound hold  for $m_{h}$. Obviously, $m_{h}$ now has to be much larger than that
in the CPV-SUSY case. For our study, this constitutes a crucial difference
between these two scenarios for a given set of masses for the gluino and the
lighter top squark.
\item If the SM is the only theory relevant for the LHC, then 
the dominant signal process is
from $pp \to t\tbar H$, where $H$ is the SM Higgs boson for which the LEP bound of 
$m_H > 114.4$ GeV is valid.
\item The SM contributions coming from $pp \to  t\tbar, t\tbar $Z$, 
t \tbar b \bbar$\footnote{We thank Manas Maity for estimating this
  background using the calculation reported in \cite{Das:2007jn}.} etc., which appear as ``common background'' 
for all the above three situations.
\end{itemize}
Note that in first three scenarios the contributing processes all involve
characteristic masses and/or couplings either in the production or in the
subsequent cascades. Thus observations made there directly carry crucial 
information on the scenario involved and hence may help discriminate the
same from the others. 

The SM contributions in the last item of the above list are
not sensitive, in any relevant way, to the details of any new physics scenarios. 
Thus they appear as universal backgrounds to the chosen signal coming from all of the 
other three scenarios.  The major sources in this category are 
(i) $t \tbar$ production with a $c$-jet from QCD 
radiation mistagged as the third $b$-jet (we assume  mistagging probability 
to be 1/25 \cite{Baer:2007ya}), (ii) $t\bar{t}b\bar{b}$ production where the semileptonic decays 
of the $t$ quarks produce the hard, isolated OSD pair and 
(iii) $t\bar{t}Z$ production where the $Z$ decays into $b$-quarks and the leptons 
come from $t$-decay.

\begin{figure}[h]
\begin{center}
\vspace*{-4.2cm}
\centerline{\epsfig{file=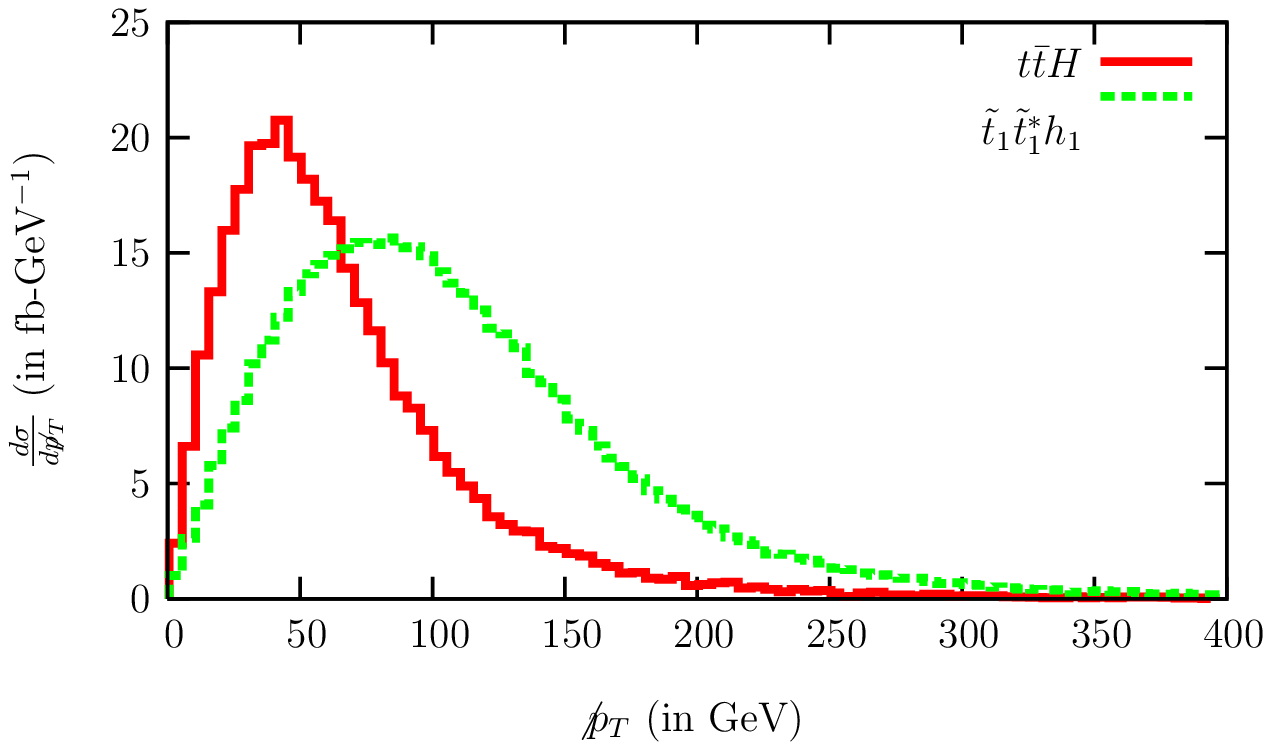,width=15.0 cm,height=30.0cm,angle=0}}
\vspace*{-20.0cm}
\vspace*{2.0cm}
\caption{$\not{p}_T$ distribution with arbitrary normalisation for the
  CPV-SUSY $\tilde{t}_1\tilde{t}_{1}^{*}h_1$ and the SM $t\bar{t}H$ background.}
\end{center}
\vspace*{-1.0cm}
\label{fig5}
\end{figure}

The most effective way to reduce the contribution from $t\bar{t}H$ in the SM(with
$m_H$=120 GeV) is found to come from the missing $p_T$ distributions. In Figure 2,
we present the $\!\!\not{p}_T$ distribution for our proposed signal, arising from
the associated lightest Higgs production along with a stop sqaurk pair. Since the plots demonstrate that the
CPX signal contains more events with $\!\!\not{p}_T$ on the higher side
(due to the massive lightest neutralino pair in the final state), an
appropriate $ \!\!\not{p}_T $-cut is clearly useful. Therefore, we have
subjected our generated events to the additional requirement 

\hspace{5cm} $\!\!\not{p}_T \ge 110~ $ GeV.

This is added to the basic cuts listed earlier, yielding an overall efficiency 
factor denoted here by $\epsilon$ which contains the effects of all
cuts described so far as well as those to be mentioned later in the text. 
The finally important numbers for the signal
and any of the faking scenarios are thus given by the quantity $\sigma\times\epsilon$,
$\sigma$ being the cross section for the aforementioned final state without
any cuts.

In case the SM is the only relevant theory for such final states at the LHC, 
$pp\rightarrow{t\bar{t}H}$ as well as the sources of `common backgrounds'
will contribute to our final state. In this, one will have to take
$m_H\geq114.4$ GeV to be consistent with the experimental observations. 
The missing-${p_T}$ cut of $\!\!\not{p}_{T} >110$ GeV effectively reduces
events of both these types. Thus having enough signal events
above the standard model predictions is ensured in this search strategy.

However, the same final state can have strong contributions from strong production
such as $pp \to \tilde{g} \tilde{g}$, followed by a cascade like 
\[ \gluino \to t \sstopc1 \to t\tbar \ntrl1 \to b \bbar W^+ W^- \ntrl1 \]
While these may add
to the signal strength, there is always the possibility that the fluctuation
in the gluino-induced events owing to the uncertainties of strong interaction
will tend to submerge the channels of our real interest, namely, the
associated production of the neutral Higgs bosons. In the same way, contributions
from strong processes may also fake the proposed signals in $CP$-conserving
SUSY. The next task, therefore, is to devise acceptance criteria to 
avoid such fake events. We take as representative the gluino pair
production process as the interfering channel, the contributions
from squarks being small at the corresponding parameter region.

\begin{figure}[t]
\begin{center}
\vspace*{-4.2cm}
\centerline{\epsfig{file=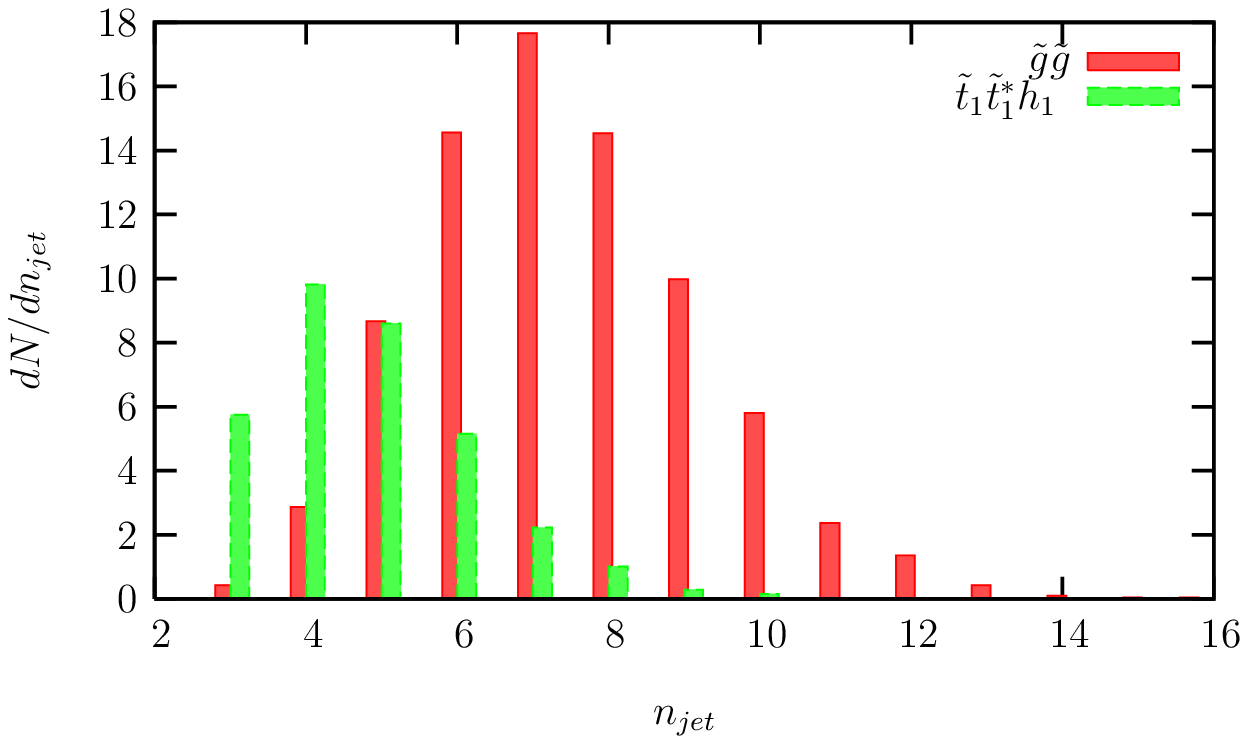,width=15.0cm,height=30.0cm,angle=0}}
\vspace*{-17.7cm}
\caption{Final state jet multiplicity distributions (with arbitrary normalisation) arising from 
$\tilde{t}_1\tilde{t}^{*}_{1}h_1$ (in green) and 
$\tilde{g}\tilde{g}$ (in red) in the CPV-SUSY scenario. }
\end{center}
\vspace*{-1.0cm}
\label{fig3}
\end{figure}

The first point to note here is that the contributions from strong 
processes leading to this final state usually have a higher jet multiplicity 
than in our case. This is evident from Figure $\ref{fig3}$ where we present the 
jet-multiplicity distribution at the CPX point. While the contributions 
from associated Higgs production peak at four jets, the 
overall peak lies at seven. This immediately suggests jet multiplicity 
as a useful acceptance criterion here, and thus we demand
$n_{j} \leq{5}$, thereby reducing considerably the artifacts of
strong processes.

There are other SUSY processes which may tend to obfuscate the presence of a
rather light Higgs boson. For example, similar final states may arise from
processes like $pp \to \sbot1 \sbotc1 h_1$, where the $\sbot1$'s decay into
a $b$-quark and the second lightest neutralino. The latter, in turn, decays 
into two leptons and the lightest supersymmetric particle (LSP). 
The number of such events, however, is 
negligible due to a highly suppressed $\sbot1$-$\sbot1$-$h_1$ coupling at  
moderate to low $\tan\beta$ values, i.e., the range of $\tan\beta$ 
answering to the CPX scenario. In case of faking in a $CP$-conserving
SUSY spectrum with high $\tan\beta$ ($\simeq 40$ or so), one has to 
study independently the $b\bar{b}$ and $\tau^+ \tau^-$ interactions, for
example, in the vector boson fusion channel \cite{Rainwater:1997dg,Rainwater:1998kj,Plehn:1999nw,Hankele:2006ma}, where the values of
the parameters can be established as different from those giving
rise to the `hole' in the CPX case.


\begin{figure}[hbt]
\vspace*{-4.2cm}
%
%
\hskip -65pt
{\epsfig{file=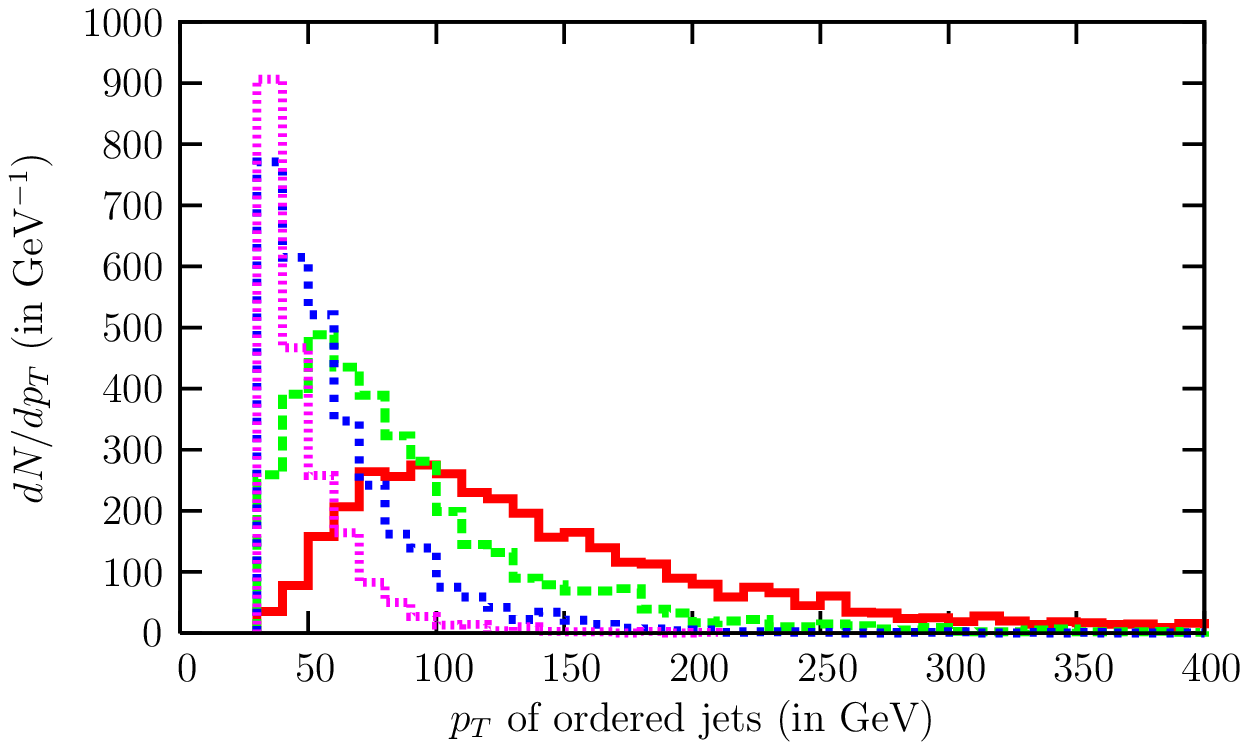,width=12.0 cm,height=26.0cm,angle=0}}
\hskip -120pt 
{\epsfig{file=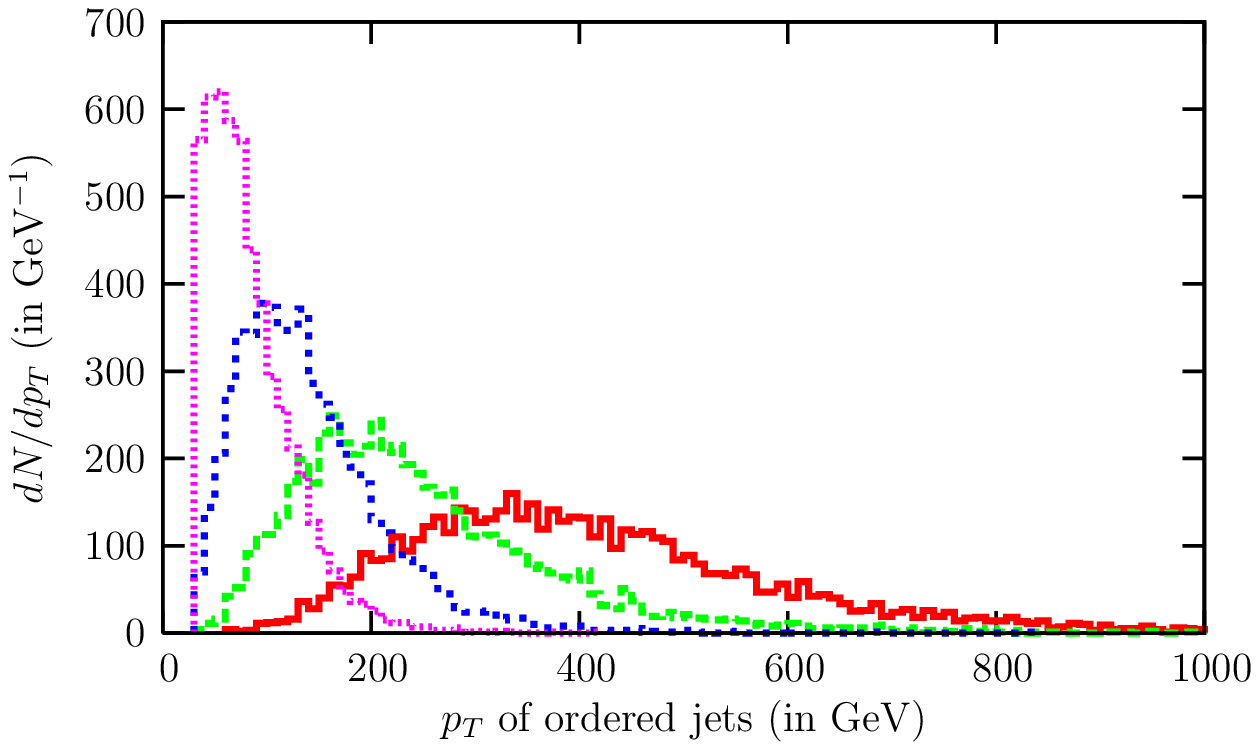,width=12.0cm,height=26.0cm,angle=0}}
\vspace*{-16cm}
\caption{Ordered $p^{jet}_T$ distributions in CPV-SUSY scenario:  
$\tilde{t}_1\tilde{t}^{*}_{1}h_1$ (left) and $\tilde{g}\tilde{g}$ (right)}
\label{fig4}
\end{figure}

The strong cascades, however, continue to remain problematic even after
imposing the jet multiplicity cut, since the production cross-sections
are  quite large and the multiplicity cut removes only about half
of the events. The next suggestion thus is to use those characteristics 
of the events  that reflect the mass (1 TeV) of the gluino in the CPX
case. The obvious distributions to look at are those of the transverse momenta
of the various jets, for the final states arising from associated Higgs
production vis-a-vis strong processes. 
It is natural to  expect that jets originating in gluino decays will have harder
$p^{jet}_{T}$ distributions compared to those coming from the associated Higgs
productions. This is obvious on comparing the left and right panels of 
Figure 4
which shows the ordered  $p_T$-distributions of jets arising
from $\tilde{t}_1\tilde{t}^{*}_{1}h_1$ and $\tilde{g}\tilde{g}$ productions
in this scenario.

Thus we further impose an upper
cut on $p^{jet}_{T}$, \emph{viz.}, $p^{jet}_{T}\leq{300}$ GeV,  which
`kills' the more energetic jets from the strong production process. Together
with the stipulated upper limit on jet multiplicity, this helps in enhancing
the share of the associated Higgs production processes in the final state
under investigation. Thus the effects of the $\!\!\not{p}_T$, multiplicity and
maximum $p^{jet}_T$ cuts all enter into the quantity $\epsilon$ determining
the final rates after all the event selection criteria are applied.

Now we are in a position to make a comparative estimate of the 
contributions to {\em dilepton + $\leq 5$ jets including three
tagged b-jets + $\!\!\not{p}_T$} from the various scenarios, and assess the
usefulness of this channel in extracting the signature of a $CP$-violating 
SUSY scenario with light neutral scalars. Such an estimate is readily
available from Tables 4 and 5.

\begin{table}[h]
\begin{center}
\begin{tabular}{||c|c|c|c|c||} \hline\hline
 & & Hard &$\sigma\times{\epsilon}$ in
fb &Final  \\
Scenarios&Processes&Cross-sections&without&number\\
&&in fb&(with)upper&at\\
& &without cut &$p^{jet}_{T}$cut &${\cal L}$=30 fb$^{-1}$ \\
\hline\hline
CPV &$pp\rightarrow \tilde{t}_1 \tilde{t}^*_1 h_1$ &
440&0.5(0.38)& 15(11)  \\

SUSY &$pp\rightarrow{t\bar{t}h_2}$ & 197&0.23(0.16)&7(5) \\
& $pp\rightarrow{t\bar{t}h_3}$ & 135 &0.23(0.17)&7(5) \\
&$pp\rightarrow{\tilde{g}\tilde{g}}$ &134 &0.70(0.167) & 21(5) \\
 \hline\hline
CPC- &$pp\rightarrow{t\bar{t}h}$ & 330 & 0.33(0.27) &10(8)  \\
SUSY &$pp\rightarrow{CPC(\tilde{g}\tilde{g})}$& 134 &0.33(0.07)& 10(2) \\\hline\hline
SM &$pp\rightarrow{SM(t\bar{t}H)}$  & 340 & 0.33(0.27) & 10(8) \\
\hline\hline
\end{tabular}
\vspace*{0.0mm}
\caption{Event rates for the CPX point, $CP$-conserving SUSY  and the
  standard model with same mass spectrum as CPX except for $m_{h,H}=117,120$ GeV for
  latter two cases respectively.}
\label{tab1}
\end{center}
\end{table}


\begin{table}[h]
\begin{center}
\vskip 10pt
\begin{tabular}{||c|c|c|c|c||} \hline\hline
 & & Hard &$\sigma\times{\epsilon}$ in
fb &Final \\
Models &Processes&Cross-sections&without&number\\
&&in fb&(with)upper&at\\
& & (without cut)&$p^{jet}_{T}$cut &${\cal L}$=30 fb$^{-1}$ \\
\hline\hline
&($pp\rightarrow{t\bar{t}}$)  & 3.7$\times 10^5$ &0.1(0.1) &3(3)  \\
Common &($pp\rightarrow{t\bar{t}Z}$)  & 370 &0.03(0.03) &1(1)  \\
Background&($pp\rightarrow{t\bar{t}b\bar{b}}$)  & 831 & 0.3(0.3) & 9(9) \\
\hline
\hline
\end{tabular}
\vspace*{0.0mm}
\caption{Event rates for the `common background' with and without the upper cut
on $p^{jet}_T$.}
\label{tab2}
\end{center}
\end{table}

Table 4 contains the contributions to the aforesaid final state
from the CPX benchmark point 1 (BP1), $CP$-conserving SUSY and a standard
model Higgs boson of masses 117 and 120 GeV respectively. These are {\em over and above the `common backgrounds'} which are listed in Table 5. In each case,
the main contributing processes and the corresponding hard 
cross-sections are shown. Also displayed are the final event rates 
once the various cuts are imposed, where the difference made by the
upper cut on $p^{jet}_T$ is clearly brought out. 

As far as the choice of parameters in $CP$-conserving SUSY is concerned,
we have used the same values of the gluino and first two generations
of squark masses as in the CPX point. It is expected that any
departure in the strong sector masses from those corresponding to the
hole in the CPX case will be found out from variables such as
the energy profile of jets, if any signal of SUSY is seen at the LHC.
Thus other regions of the MSSM parameter space are unlikely to
fake the signals of $CP$-violating situation. The value of $\tan\beta$ is also 
kept at the region allowed by the CPX hole, and any departure from this region
in a faking MSSM scenario has to show up in the branching ratios for
$h_1$ $\to$ $b{\bar{b}}$, $\tau^+ \tau^-$, using the supplementary data on
the vector boson fusion channel. Finally, although some difference from the
rates shown in Table 4 for $CP$-conserving SUSY can in principle occur due
to different values of the lighter stop mass, the overall rates are
not significantly different, so long as stop squark decays dominantly into 
either $b\chi^+_1$ or $t\chi^0_1$. Thus the choice of the $CP$-conserving
SUSY parameters in Table 4 can be taken as representative. We checked that for
smaller choice of $\tilde{t}_1$ mass also and the number is still smaller than
CPX contribution.

It is easy to draw one's own conclusion from these two tables
about the viability of the suggested search strategy.
With the selection criteria  proposed in this
paper (without the upper cut on jet $p_T$) the size of the signal (50 events)
from the dominant processes in CPV-SUSY for only
30 fb$^{-1}$ of integrated luminosity easily dwarfs the common SM background (13 events).
Moreover, the signal  size is much larger than that in  the CPC scenario
(with comparable squark and gluino masses) or in the SM.
Thus, important hints regarding the existence of new physics and its nature
will be available  at this stage (we assume that the gluino mass and some
other important  parameters will be determined from complimentary
experiments). The presence of the lightest Higgs boson and its not so heavy mates
becomes clear after the upper cut on $p_T$ since nearly 75\% of the new physics events are
now induced by them. Clearly, even after imposing the upper cut on $p^{jet}_T$, 
the signals can rise above the SM backgrounds at more than
5$\sigma$ level within a moderate integrated luminosity like
30fb$^{-1}$. This can be further magnified with the accumulation of luminosity.
On the other hand, it is not too optimistic to assume that important hints will
be available with only 10 fb$^{-1}$ of integrated luminosity.

\begin{table}[hbt]
\renewcommand\baselinestretch{0.2}
\begin{center}
\begin{tabular}{||c|c|c|c|c|c|c||}
\hline\hline
Parameters&BP2&BP3&BP4&BP5&BP6&BP7\\
\hline\hline
$\tan{\beta}$&5.0&7.0&7.0&6.0&5.0&7.0\\
\hline
$m_{h_1}$ (GeV) &40.5&40.4&49.0&45.1&30.0&30.0\\
\hline\hline
\end{tabular}
\vspace*{0.0mm}
\caption{Benchmark points within the LEP-hole in $m_{h_1}$-$\tan\beta$ plane.}
\label{tab6}
\end{center}
\end{table}

Before we end this discussion, we show the viability of this signal in 
other regions of the CPX hole. It has already been noted in the literature that
the size and the exact location of the hole in the parameter space
depend on the method of calculating the loop corrections \cite{Lee:2003nta,Frank:2006yh,Hahn:2006np}.
However, the  calculations agree qualitatively and confirm the presence of the
hole. To be specific we have chosen points from  the hole
as presented by \cite{Bechtle:2006iw}.

In Table $\ref{tab6}$ we present different sets of values of $\tan\beta$ and 
$m_{h_1}$, keeping the other parameters fixed at their CPX values. 
These correspond to six different regions of the LEP hole and are
termed as benchmark points 2 -7 (BP2 - BP7), all 
within the hole. 
The analysis for each of these points is an exact parallel 
of that already presented for the first benchmark point. 
We have computed the generic sensitivity of LHC to the `hole'
corresponding to each of these benchmark points,
the results being summarised in Table $\ref{tab7}$.
It is clear from this Table that we always have enough events 
($> 15$) in our attempt to probe the 
LEP-hole even with an integrated luminosity of 30 fb$^{-1}$.
As the luminosity accumulates a statistically significant signal will be
obtainable from any corner of this hole.

\vskip 5pt
\begin{table}[hbt]
\renewcommand\baselinestretch{1.1}
\begin{center}
{\normalsize
\begin{tabular}{||c|c|c|c|c|c||} \hline\hline
Bench& & Cross&$\sigma\times{\epsilon}$ in
fb& Total& Events \\
\vspace{-0.2cm}
Marking&Processes&-section&with&$\sigma\times{\epsilon}$&at\\
\vspace{-0.2cm}
points&&in fb&upper&in fb& ${\cal L}$=30 \\
\vspace{-0.2cm}
& & &$p^{jet}_{T}$cut &&fb$^{-1}$\\
\hline\hline
&$pp\rightarrow{\tilde{t}_{1}\tilde{t}^{*}_{1}}h_1$  & 560 &0.47 &&  \\
BP2&$pp\rightarrow{t\bar{t}h_2}$ & 180 &0.10 &0.67&20  \\
&$pp\rightarrow{t\bar{t}h_3}$ & 145 & 0.10 &&   \\
\hline
\vspace{-0.2cm}
&$pp\rightarrow{\tilde{t}_{1}\tilde{t}^{*}_{1}h_1}$  & 437 &0.37 &&  \\
BP3&$pp\rightarrow{t\bar{t}h_2}$ & 180 &0.10 &0.63&19  \\
&$pp\rightarrow{t\bar{t}h_3}$ & 195 & 0.16 &&   \\
\hline
\vspace{-0.2cm}
&$pp\rightarrow{\tilde{t}_{1}\tilde{t}^{*}_{1}h_1}$  & 350 &0.34 &&  \\
BP4&$pp\rightarrow{t\bar{t}h_2}$ & 135 &0.10 &0.60&18  \\
&$pp\rightarrow{t\bar{t}h_3}$ & 178 & 0.16 && \\
\hline
\vspace{-0.2cm}
&$pp\rightarrow{\tilde{t}_{1}\tilde{t}^{*}_{1}h_1}$  & 422 &0.37 &&  \\
BP5&$pp\rightarrow{t\bar{t}h_2}$ & 154 &0.13 &0.69& 21  \\
&$pp\rightarrow{t\bar{t}h_3}$ & 167 & 0.19 &&   \\
\hline
\vspace{-0.2cm}
&$pp\rightarrow{\tilde{t}_{1}\tilde{t}^{*}_{1}h_1}$  & 760 &0.59 &&  \\
BP6&$pp\rightarrow{t\bar{t}h_2}$ & 170 &0.11 &0.88& 26  \\
&$pp\rightarrow{t\bar{t}h_3}$ & 170 & 0.18 &&  \\
\hline
\vspace{-0.2cm}
&$pp\rightarrow{\tilde{t}_{1}\tilde{t}^{*}_{1}h_1}$  & 590 &0.48 &&  \\
BP7&$pp\rightarrow{t\bar{t}h_2}$ & 100 &0.06 & 0.74 & 22  \\
&$pp\rightarrow{t\bar{t}h_3}$ & 210 & 0.20 &&  \\
\hline
\hline
\end{tabular}
}
\vspace*{0.0mm}
\caption{Final numbers of signal events for 30 fb$^{-1}$ integrated 
luminosity at various benchmark points in the LEP hole. }

\label{tab7}
\end{center}
\end{table}

\section{Summary and Conclusions}

Taking a cue from the frequently discussed possibility of 
$CP$-violation in MSSM and its phenomenological
consequences at colliders, we explore a popular benchmark scenario
(called the CPX scenario) of this broad framework. The study is motivated 
by recent analyses which reveal that the LEP, in its standard Higgs searches,
could not probe some of  the region in the parameter space of this scenario
having low $m_{h_1}$ and low to moderate 
$\tan\beta$ values. We concentrated on this `unfilled hole' in the parameter space 
and studied how well  LHC could explore it. 

We have found that the associated production of the lightest Higgs boson 
(which may evade the LEP bound and be as light as 50 GeV or smaller) and two
of its `light' mates along with a pair of top quarks and top squarks could be 
extremely useful in reaching out to this region. This is because one can now
exploit modes where the involved couplings and the masses are very 
characteristic of the $CP$-violating SUSY scenario. 
The particular signal we choose for the study is
3\emph{-tagged $b$-jets + dilepton + tagged 
jets + missing transverse momentum}, the total number of jets
being within 5. It is shown that the entire `LEP-hole' can be
probed in detail in this final state with less than 
50 fb$^{-1}$ of LHC data, and that
the $CP$-violating SUSY effects cannot be faked
even by a combined effect from the contending scenarios like 
$CP$-conserving MSSM and/or the standard model.

\vskip 15pt
\noindent
{\bf Acknowledgments:} We thank Siba Prasad Das for help in the initial stages of
simulation and Manas Maity for providing
some important information on the calculation of the backgrounds. 
We also thank Subhaditya Bhattacharya, Sudhir K. Gupta, Sujoy Poddar, 
Alexander Pukhov and Gaurab Sarangi for helpful discussions and suggestions 
on the code. 
AD thanks Apostolos Pilaftsis for a useful private communication. 
PB, AKD and BM thank the Theoretical Physics Group of Indian
Association for the Cultivation of Science, Kolkata, India for 
hospitality while the project was in progress. AD acknowledges the hospitality
of Regional Centre for Accelerator-based Particle Physics (RECAPP),
Harish-Chandra Research Institute during the latter part of the project. 
Computational work for this study was partially carried out in the cluster
computing facility at Harish-Chandra Research Institute (HRI)
(http://cluster.mri.ernet.in). This work is partially supported by the RECAPP, Harish-Chandra Research Institute, and funded by the Department of Atomic Energy, Government of India under the XIth 5-year Plan. AD's work was supported by
DST, India, project no SR/S2/HEP-18/2003.


\newpage
\begin{thebibliography}{99}

\bibitem{Barate:2003sz}
  R.~Barate {\it et al.}  [LEP Working Group for Higgs boson searches],
  Phys.\ Lett.\  B {\bf 565}, 61 (2003)
  [arXiv:hep-ex/0306033]; \\
see also 
$http://lephiggs.web.cern.ch/LEPHIGGS/www/Welcome.html$

\bibitem{Schael:2006cr}
  S. Schael {\it et al.} [ALEPH Collaboration],
  Eur. Phys. J. C {\bf 47}, 547 (2006) 
  [arXiv:hep-ex/0602042]; \\
see also $http://lephiggs.web.cern.ch/LEPHIGGS/www/Welcome.html$

\bibitem{Pilaftsis:1998pe}
  A.~Pilaftsis,
  Phys.\ Rev.\  D {\bf 58}, 096010 (1998)
  [arXiv:hep-ph/9803297].

\bibitem{Pilaftsis:1998dd}
  A.~Pilaftsis,
  Phys.\ Lett.\  B {\bf 435}, 88 (1998)
  [arXiv:hep-ph/9805373].

\bibitem{Carena:2000ks}
  M.~S.~Carena, J.~R.~Ellis, A.~Pilaftsis and C.~E.~M.~Wagner,
  Phys.\ Lett.\  B {\bf 495}, 155 (2000)
  [arXiv:hep-ph/0009212].


\bibitem{Bechtle:2006iw}
  P.~Bechtle  [LEP Collaboration],
  PoS {\bf HEP2005}, 325 (2006)
  [arXiv:hep-ex/0602046].



\bibitem{Carena:2002bb}
  M.~S.~Carena, J.~R.~Ellis, S.~Mrenna, A.~Pilaftsis and C.~E.~M.~Wagner,
  Nucl.\ Phys.\  B {\bf 659}, 145 (2003)
  [arXiv:hep-ph/0211467].
\bibitem{Accomando:2006ga}
 See, for example,  E.~Accomando {\it et al.},
  [arXiv:hep-ph/0608079], p109.


\bibitem{Li:2006hq}
  Z.~Li, C.~S.~Li and Q.~Li,
  Phys.\ Rev.\  D {\bf 73}, 077701 (2006)
  [arXiv:hep-ph/0601148].

\bibitem{Demir:1999hj}
  D.~A.~Demir,
  Phys.\ Rev.\  D {\bf 60}, 055006 (1999)
  [arXiv:hep-ph/9901389].

\bibitem{Pilaftsis:1999qt}
  A.~Pilaftsis and C.~E.~M.~Wagner,
  Nucl.\ Phys.\  B {\bf 553}, 3 (1999)
  [arXiv:hep-ph/9902371].

\bibitem{Dedes:1999sj}
  A.~Dedes and S.~Moretti,
  Phys.\ Rev.\ Lett.\  {\bf 84}, 22 (2000)
  [arXiv:hep-ph/9908516].

\bibitem{Dedes:1999zh}
  A.~Dedes and S.~Moretti,
  Nucl.\ Phys.\  B {\bf 576}, 29 (2000)
  [arXiv:hep-ph/9909418].
\bibitem{Choi:1999aj}
  S.~Y.~Choi and J.~S.~Lee,
  Phys.\ Rev.\  D {\bf 61}, 115002 (2000)
  [arXiv:hep-ph/9910557].




\bibitem{Choi:2000wz}
  S.~Y.~Choi, M.~Drees and J.~S.~Lee,
  Phys.\ Lett.\  B {\bf 481}, 57 (2000)
  [arXiv:hep-ph/0002287].


\bibitem{Kane:2000aq}
  G.~L.~Kane and L.~T.~Wang,
  Phys.\ Lett.\  B {\bf 488}, 383 (2000)
  [arXiv:hep-ph/0003198].
\bibitem{Choi:2001pg}
  S.~Y.~Choi, K.~Hagiwara and J.~S.~Lee,
  Phys.\ Rev.\  D {\bf 64}, 032004 (2001)
  [arXiv:hep-ph/0103294].
\bibitem{Heinemeyer:2001qd}
  S.~Heinemeyer,
  Eur.\ Phys.\ J.\  C {\bf 22}, 521 (2001)
  [arXiv:hep-ph/0108059].

\bibitem{Choi:2001iu}
  S.~Y.~Choi, K.~Hagiwara and J.~S.~Lee,
  Phys.\ Lett.\  B {\bf 529}, 212 (2002)
  [arXiv:hep-ph/0110138].
\bibitem{Arhrib:2001pg}
  A.~Arhrib, D.~K.~Ghosh and O.~C.~W.~Kong,
  Phys.\ Lett.\  B {\bf 537}, 217 (2002)
  [arXiv:hep-ph/0112039].

\bibitem{Ibrahim:2002zk}
  T.~Ibrahim and P.~Nath,
  Phys.\ Rev.\  D {\bf 66}, 015005 (2002)
  [arXiv:hep-ph/0204092].
\bibitem{Choi:2002zp}
  S.~Y.~Choi, M.~Drees, J.~S.~Lee and J.~Song,
  Eur.\ Phys.\ J.\  C {\bf 25}, 307 (2002)
  [arXiv:hep-ph/0204200].



\bibitem{Ham:2002ps}
  S.~W.~Ham, S.~K.~Oh, E.~J.~Yoo, C.~M.~Kim and D.~Son,
  Phys.\ Rev.\  D {\bf 68}, 055003 (2003)
  [arXiv:hep-ph/0205244].





\bibitem{Akeroyd:2003jp}
  A.~G.~Akeroyd,
  Phys.\ Rev.\  D {\bf 68}, 077701 (2003)
  [arXiv:hep-ph/0306045].



\bibitem{Ghosh:2004cc}
  D.~K.~Ghosh, R.~M.~Godbole and D.~P.~Roy,
  Phys.\ Lett.\  B {\bf 628}, 131 (2005)
  [arXiv:hep-ph/0412193].

\bibitem{Ghosh:2004wr}
  D.~K.~Ghosh and S.~Moretti,
  Eur.\ Phys.\ J.\  C {\bf 42}, 341 (2005)
  [arXiv:hep-ph/0412365].


\bibitem{PvtApost}
  A. Pilaftsis, private communication.


\bibitem{Pukhov:2004ca}
  A.~Pukhov,
``CalcHEP 3.2: MSSM, structure functions, event generation, batchs, and
generation of matrix elements for other packages'',
  [arXiv:hep-ph/0412191].

\bibitem{Ellis:2006eh}
  J.~R.~Ellis, J.~S.~Lee and A.~Pilaftsis,
  Mod.\ Phys.\ Lett.\  A {\bf 21}, 1405 (2006)
  [arXiv:hep-ph/0605288].

\bibitem{Lee:2003nta}
  J.~S.~Lee, A.~Pilaftsis, M.~S.~Carena, S.~Y.~Choi, M.~Drees, J.~R.~Ellis and C.~E.~M.~Wagner,
  Comput.\ Phys.\ Commun.\  {\bf 156}, 283 (2004)
  [arXiv:hep-ph/0307377].


\bibitem{alex}
See \emph{``http://hep.pa.msu.edu/people/belyaev/public/calchep/index.html''}.

\bibitem{Skands:2003cj}
  P.~Skands {\it et al.},
  JHEP {\bf 0407}, 036 (2004)
  [arXiv:hep-ph/0311123]; \\ see also 
$http://home.fnal.gov/\tilde{}skands/slha/$






\bibitem{Sjostrand:2001yu}
  T.~Sjostrand, L.~Lonnblad and S.~Mrenna,
  [arXiv:hep-ph/0108264].















\bibitem{Lai:1999wy}
  H.~L.~Lai {\it et al.}  [CTEQ Collaboration],
  Eur.\ Phys.\ J.\ C {\bf 12}, 375 (2000)
  [arXiv:hep-ph/9903282].

\bibitem{Pumplin:2002vw}
  J.~Pumplin, D.~R.~Stump, J.~Huston, H.~L.~Lai, P.~Nadolsky and W.~K.~Tung,
  JHEP {\bf 0207}, 012 (2002)
  [arXiv:hep-ph/0201195].

\bibitem{Das:2007jn}
  S.~P.~Das, A.~Datta, M.~Guchait, M.~Maity and S.~Mukherjee,
  arXiv:0708.2048 [hep-ph].

\bibitem{Baer:2007ya}
 See, for example, H.~Baer, V.~Barger, G.~Shaughnessy, H.~Summy and L.~t.~Wang,
  Phys.\ Rev.\  D {\bf 75}, 095010 (2007)
  [arXiv:hep-ph/0703289].





\bibitem{Rainwater:1997dg}
  D.~L.~Rainwater and D.~Zeppenfeld,
  JHEP {\bf 9712}, 005 (1997)
  [arXiv:hep-ph/9712271].
\bibitem{Rainwater:1998kj}
  D.~L.~Rainwater, D.~Zeppenfeld and K.~Hagiwara,
  Phys.\ Rev.\  D {\bf 59}, 014037 (1999)
  [arXiv:hep-ph/9808468].

\bibitem{Plehn:1999nw}
  T.~Plehn, D.~L.~Rainwater and D.~Zeppenfeld,
  Phys.\ Lett.\  B {\bf 454}, 297 (1999)
  [arXiv:hep-ph/9902434].
\bibitem{Hankele:2006ma}
  V.~Hankele, G.~Klamke, D.~Zeppenfeld and T.~Figy,
  Phys.\ Rev.\  D {\bf 74}, 095001 (2006)
  [arXiv:hep-ph/0609075].

\bibitem{Frank:2006yh}
  M.~Frank, T.~Hahn, S.~Heinemeyer, W.~Hollik, H.~Rzehak and G.~Weiglein,
  JHEP {\bf 0702}, 047 (2007)
  [arXiv:hep-ph/0611326].
\bibitem{Hahn:2006np}
  T.~Hahn, S.~Heinemeyer, W.~Hollik, H.~Rzehak, G.~Weiglein and K.~Williams,
  [arXiv:hep-ph/0611373].

\end{thebibliography}
\end{document}